\documentclass[a4paper,11pt]{article}
\usepackage{pos}

\title{2+1 flavor fine lattice simulation at finite temperature with domain-wall fermions}
\ShortTitle{$N_f=2+1$, $T>0$ fine lattice simulation with DWF}

\author[a]{Sinya Aoki}
\author*[b]{Yasumichi Aoki}
\author[c]{Hidenori Fukaya}
\author[d,e]{Shoji Hashimoto}
\author[a]{Issaku~Kanamori}
\author[d,e]{Takashi Kaneko}
\author[a]{Yoshifumi Nakamura}

\affiliation[a]{Center for Gravitational Physics, Yukawa Institute for Theoretical Physics, Kyoto University, Kyoto 606-8502, Japan}
\affiliation[b]{RIKEN Center for Computational Science, 
  Kobe 650-0047, Japan}
\affiliation[c]{Department of Physics, Osaka University, Toyonaka, Osaka 560-0043, Japan}
\affiliation[d]{High Energy Accelerator Organization (KEK), Tsukuba 305-0801, Japan}
\affiliation[e]{School of High Energy Accelerator Science, SOKENDAI (The Graduate University for Advanced Studies), Tsukuba 305-0801, Japan}

\emailAdd{yasumichi.aoki@riken.jp}

\abstract{
Simulations for the thermodynamics of the 2+1 flavor QCD are performed employing chiral fermions. The use of M\"obius domain-wall fermions with stout-link smearing is more effective on the finer lattices where 
all the relevant chiral symmetries are realized more accurately.
We report on the initial simulations near the (pseudo) critical point using the line of constant physics with an average $ud$ quark mass slightly heavier than physical at $a\lesssim 0.1$ fm.}

\FullConference{%
 The 38th International Symposium on Lattice Field Theory, LATTICE2021
  26th-30th July, 2021
  Zoom/Gather@Massachusetts Institute of Technology
}

\begin{document}
\begin{flushright}
YITP-21-151
\end{flushright}
\maketitle

\section{Introduction}

Nature of the phase transition of $2+1$-flavor QCD is a prime interest
as it provides a good approximation of what happens in the real world
when the temperature gets high enough to liberate the quark degree of
freedom. Especially the ``physical point'' simulations, where the
degenerate $ud$ quark mass is set as their average physical value and strange
quark mass is tuned to physical, are useful and have direct relevance 
for the understanding of the history of the universe and
physics in the heavy ion/high energy experiments.
These days there have emerged much interests in investigating, not only
the physical point, but also extended region like fictitious $ud$ chiral 
limit.  More in general, the phase structure of the Columbia plot
is getting match attention
%\footenote{See for example recent reports 
\cite{
Philipsen:2019rjq, % Lattice 2019 review
Guenther:Lat2021,  % Lattice 2021 review
Lahiri:Lat2021}.   % Lattice 2021 topical

Here we report on the newly started project aiming to study the thermodynamics
near the physical point in the Columbia plot using a chiral fermion formulation.
There was a study by HotQCD collaboration using domain-wall fermions at
a course lattice (with the temporal extent $N_t=8$) \cite{Bazavov:2012qja}.
In the studies by JLQCD for $2$-flavor ($N_f=2$) QCD 
\cite{Tomiya:2016jwr,Aoki:2020noz,Aoki:2021qws}
it is shown that the fine lattice simulations are indispensable to
maintain the underlining symmetries $SU(2)$ and $U(1)$ chiral,
especially to study the fate of $U(1)_A$. To understand the phase structure
near the $ud$ chiral limit it would be important have good control of these
symmetries as the fate of the symmetries could change the order of the
transition \cite{Pisarski:1984ms}.
This made us plan the $N_f=2+1$ flavor simulations using the same setup
as $N_f=2$, which uses M\"obius domain-wall fermions 
\cite{Brower:2012vk} (scale-factor-2 Shamir-type fermions) 
with the stout-link smearing in the Wilson kernel.
Some $N_f=2+1$ studies have already been done following the same strategy as 
$N_f=2$ where the gauge coupling is fixed and the ``ud'' mass is changed.
The preliminary
results have been reported in this conference \cite{Suzuki:Lat2021}.
% \footnote{Parallel talk given by K.~Suzuki.}
Yet another project using the same action, 
but with a line of constant physics to change only the temperature, 
has been started.
 
In this report, our main focus is to discuss how we obtain and use the
line of constant physics of the action we use for the $N_f=2+1$ QCD.
In these studies we fix the domain-wall fermion parameters:
the domain-wall height as $M_5=1$ as usual in JLQCD and 
the 5th dimension size as $L_s=12$.
After discussing these setup, some early and preliminary results are
shown.

\section{Scale setting and line of constant physics}

We would like to express the lattice spacing and physical strange quark mass
point as functions of the gauge coupling $\beta$:
$a(\beta)$ and $m_s(\beta)$, where the latter is the strange quark
mass in the lattice action. 
The ratio of strange and average $ud$ quark mass is known and we can 
obtain the lattice $m_{ud}(\beta)$ using the ratio.
% See for example the latest FLAG review.

\subsection{Lattice spacing}
\label{sec:lattice_spacing}

\begin{figure}
\begin{center}
 \includegraphics[width=7.3cm]{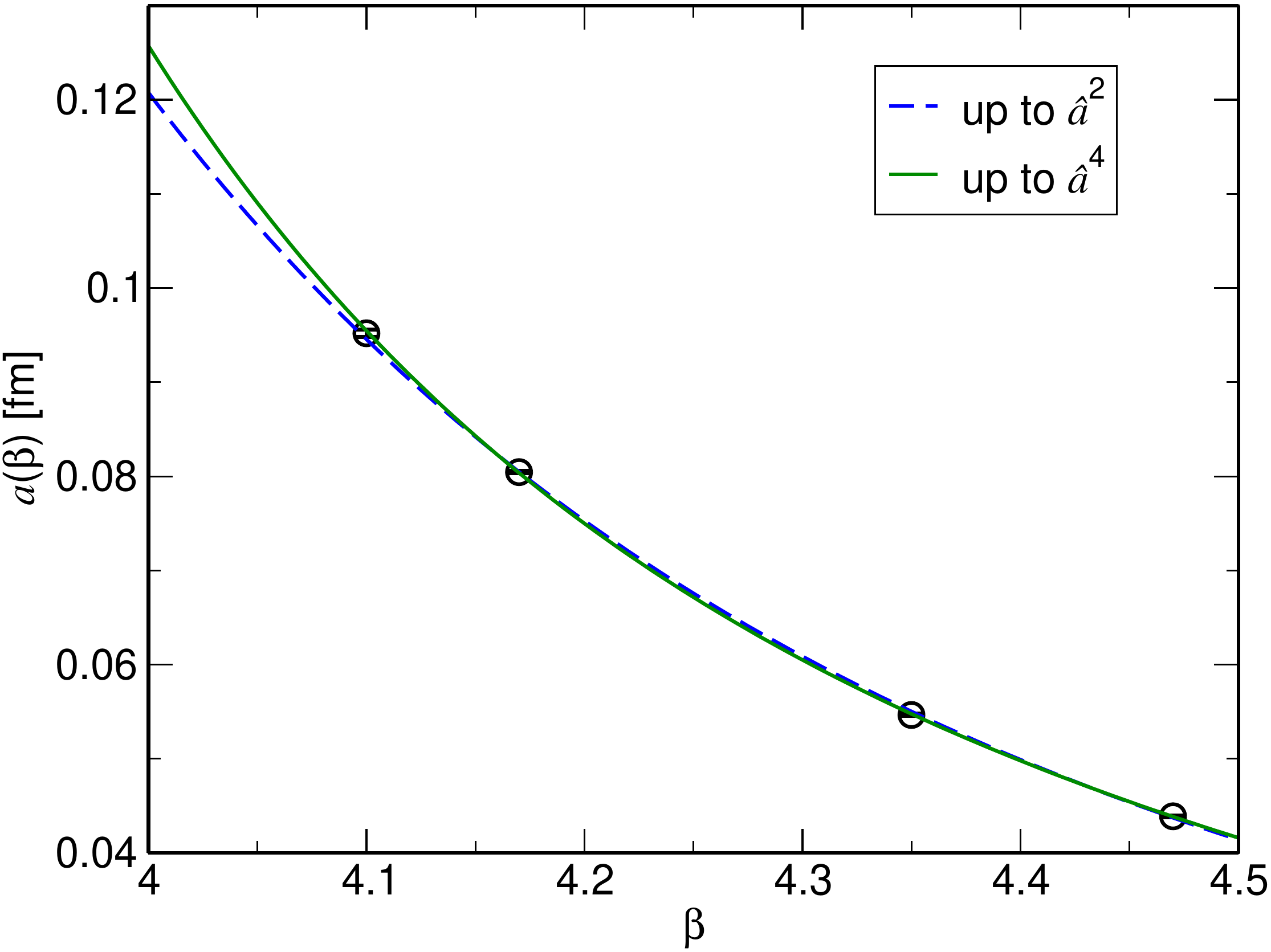}
 \hspace{6pt}
 \includegraphics[width=6.9cm]{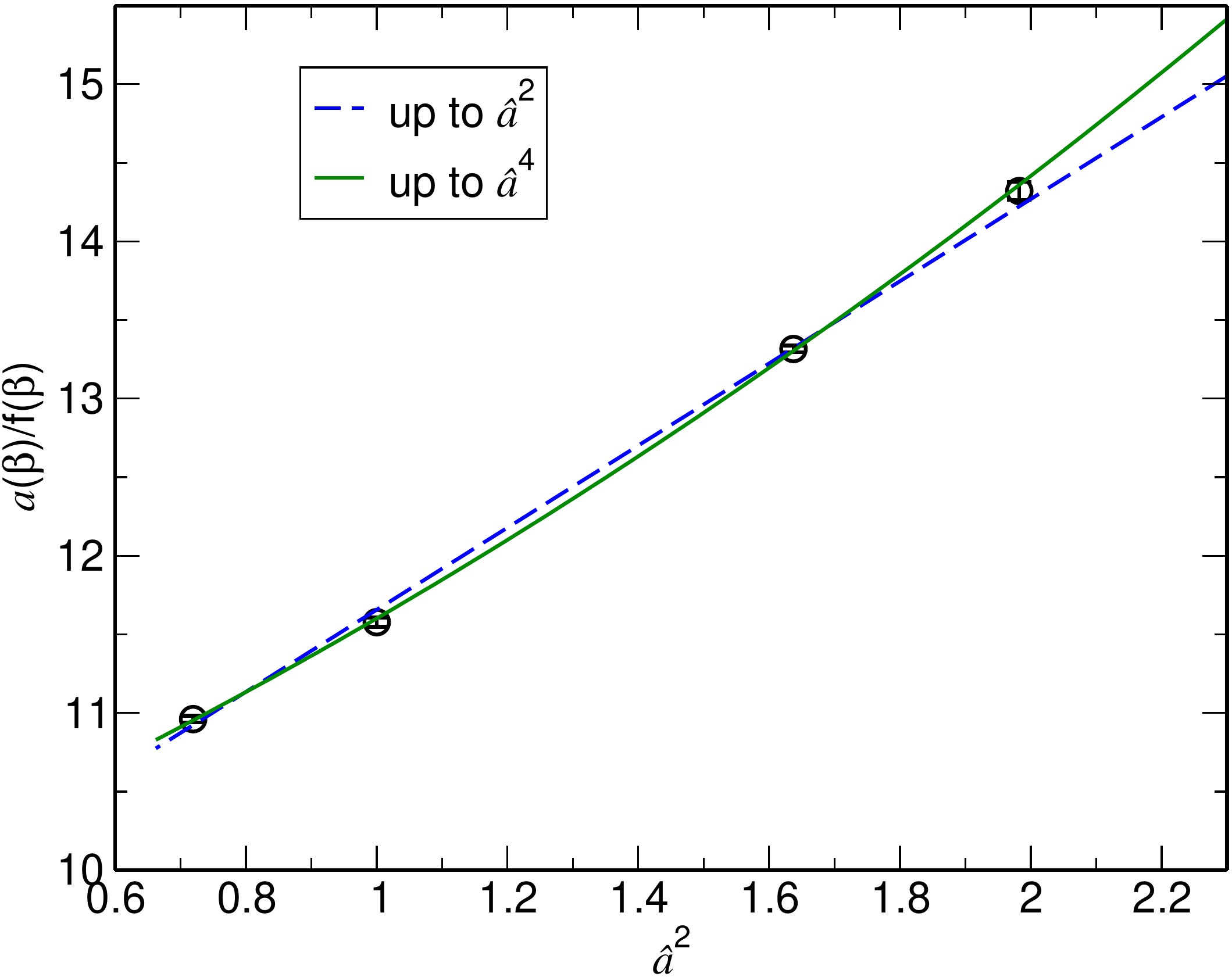}
\end{center}
 \caption{Lattice spacing as a function of $\beta$ (left) and 
 lattice spacing divided by two-loop scaling as a function of an effective 
 lattice spacing squared (right).}
 \label{fig:a-b}
\end{figure}

We restrict the lattice spacing to be used in this finite temperature
study not to be far away from the region used in the zero temperature
JLQCD studies.  These include the coarsest lattices generated only for an
unpublished pilot study $a\simeq 0.095$ fm, as well as finer lattices:
 $a\simeq 0.08$, $0.055$, $0.044$ fm. 
These lattice spacings are determined using $t_0$.
% (see Sec.~\ref{app:scale} for details).
In this way we can avoid performing expensive extra zero temperature simulations
as much as possible.
The left panel of Fig.~\ref{fig:a-b} shows the results of the lattice spacing.

There is a method often used to parameterize the lattice spacing 
as a function of the gauge coupling using two-loop beta function
with correction terms 
proposed by Edwards et al \cite{Edwards:1997xf},
\begin{equation}
% a = c_0 f(g^2)(1+c_2\hat{a}(g)^2), 
%  \label{eq:a_scale}
a = c_0 f(g^2)(1+c_2\hat{a}(g)^2 + c_4\hat{a}(g)^4).
\label{eq:a_scale2}
\end{equation}
where
\begin{eqnarray}
% \hat{a}(g)^2 & \equiv & [f(g^2)/f(g_0^2)]^2,\label{eq:ahat}\\
 \hat{a}(g) & \equiv & \frac{f(g^2)}{f(g_0^2)},\label{eq:ahat}\\
 f(g^2) & \equiv & (b_0g^2)^{-b_1/2b_0^2}\exp\left(-\frac{1}{2b_0 g^2}\right),
  \\%\nonumber\\
 b_0&=&\frac{1}{(4\pi)^2}\left(11-\frac{2}{3}N_f\right),\;\;\;
  b_1=\frac{1}{(4\pi)^4}\left(102-\frac{38N_f}{3}\right),
\end{eqnarray}
$g^2=6/\beta$, $N_f=3$,
$c_0$, $c_2$ and $c_4$ are free parameters of the fit.
We set the reference gauge coupling $g_0$
from the $\beta$ value of the second finest lattice $g_0^2=6/4.35$.
$f(g^2)$ expresses the scaling from the two-loop beta function,
which is scheme independent. Beyond two loop, 
scheme dependence appears and is not convenient for this purpose.
The $c_2$ and $c_4$ terms are meant to absorb the lattice discretization error.
But, in practice they are also playing the role of absorbing the
remnant RG scaling beyond two loop,
which can be seen in the right panel of Fig.~\ref{fig:a-b}, where
$a/f(g^2)$ is plotted as a function of $\hat{a}^2$. 
The variation from 11 to 14 is too large
to be regarded as a discretization error for domain-wall fermions at $a<0.1$ fm.
Apart from the role of each terms one wants to check the effectiveness
of the formula by looking at this figure.
If the linearity is good one does not have to include $\hat{a}^4$ term in Eq.~\ref{eq:a_scale2} to parameterize the lattice spacing. 
While it turns out that the linearity is marginally good,
the fit (shown as dashed line) results in large $\chi^2/dof=6.6$. 
Therefore, we shall adopt the parameterization using up to $\hat{a}^4$
(shown as solid line) which gives $\chi^2/dof=1.6$. 
Resulting $a(\beta)$ parameterizations have been shown in Fig.~\ref{fig:a-b}.

The simulation range for our coarser lattice include $\beta=4.0$ as 
the lower edge. The $a(\beta)$ there is an extrapolation and has
up to a few percent systematic uncertainty, estimated through
an experiment of omitting the coarsest data to check how the two fits work
at the coarsest data point.

\subsection{Quark mass}

The values of strange and average up, down quark masses are known to a good precision.
To obtain the line of constant physics given the parameterization
of the lattice spacing $a(\beta)$ in the previous subsection, 
we use the strange quark mass input using the relation,
\begin{equation}
 m_q^R = Z_m m_q^{latt} \cdot a^{-1}(\beta),
  \label{eq:quark_mass_relation}
\end{equation}
where $m_q^R$ is the dimension-full renormalized quark mass of flavor $q$.
We shall use $\overline{\mbox{MS}}$ scheme at the renormalization scale
$\mu=2$ GeV for the renormalization constant $Z_m$. 
Once a parameterization of the quark mass renormalization factor
$Z_m(\beta)$ is obtained, $m_q^{latt}(\beta)$ (multiplicatively renormalizable
mass in lattice units) may be computed.
This is a method alternative to the commonly-used hadron-mass input.

We shall use the following numbers for the quark masses for the $N_f=2+1$
physical point:
\begin{eqnarray}
 m_s^R & = & 92\; \mbox{MeV}\\ \label{eq:m_s}
 m_s^R/m_{ud}^R & = & 27.4, \label{eq:m_ud}
\end{eqnarray}
based on the FLAG2019 averages \cite{Aoki:2019cca}: 
$m_s^R = 92.0(1.1)$ and $m_s^R/m_{ud}^R=27.42(12)$.

To determine the bare quark mass $m_q$ in the domain-wall fermion action, 
the residual mass $m_{res}$ due to a finite 5th dimension
needs to be subtracted,
\begin{equation}
% m_q^{latt} = m_q^{bare} + m_{res},
 m_q^{latt} = m_q + m_{res}.
  \label{eq:m_q_latt}
\end{equation}
For $\beta\ge 4.17$
\footnote{$\beta=4.17$ is expected to be in the transition region of 
the $N_t=16$ lattice at physical quark masses.}
the residual mass is $m_{res}\lesssim 1$ MeV,
which is about the same size of the error in $m_s^R$. Therefore
we can safely neglect the effect of $m_{res}$ for the strange quark mass there.
The physical $ud$ quark mass is larger than the residual mass, 
$m_{ud}^{R}>m_{res}$. However, the size is comparable as we will see below.

\begin{figure}
\begin{center}
 \includegraphics[width=7cm]{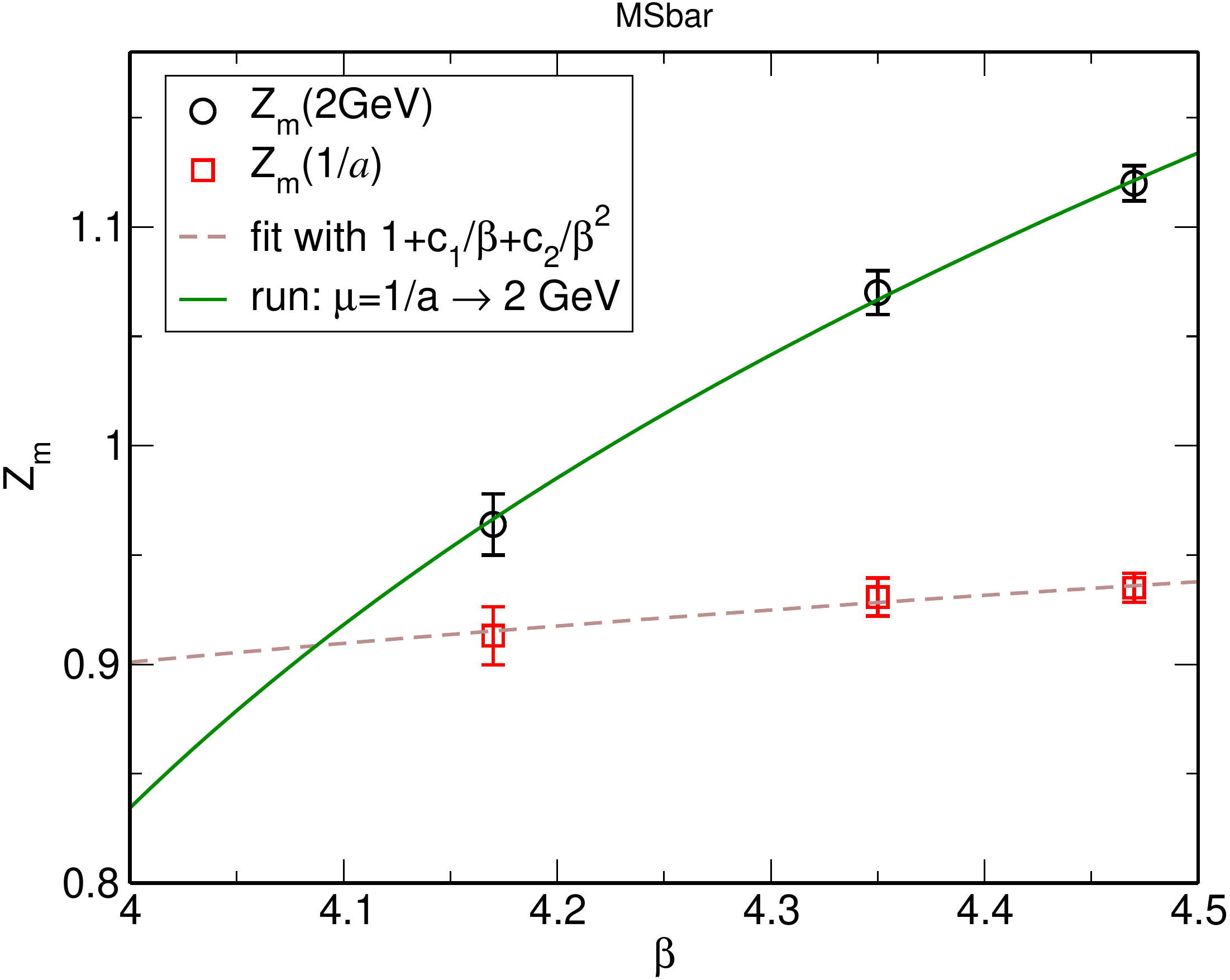}
\end{center}
 \caption{
 Renormalization factor plotted as a function of $\beta$ at $2$ GeV (black) and that run to $\mu=a^{-1}$
 using $\overline{\mbox{MS}}$ NNNLO running.
 Interpolation for the red squares is done through the fit with 
 second-order polynomial in $g^2\propto\beta^{-1}$. 
 Then it is run again to $\mu=2$ GeV using the $a(\beta)$ parameterization
 with $O(\hat{a}^4)$.}
 \label{fig:Zm-beta}
\end{figure}

We use $Z_m$ obtained for the three finer lattice spacings 
\cite{Tomii:2016xiv} and parameterize that as a smooth function
of $\beta$. Fig.~\ref{fig:Zm-beta} shows the measured $Z_m$ and its
parameterization $Z_m(\beta)$ using a method described below.
% which can be extended to a bit of extrapolation to $\beta$ values we may need.
%

Let us first determine $Z_m$ at the scale $\mu=a^{-1}$ run from $\mu=2$ GeV,
expecting the large log effect ($\log(a\mu)$) is removed.
The RG running is performed using NNNLO \cite{Chetyrkin:1999pq} 
in $\overline{\mbox{MS}}$ scheme.
Resulting $Z_m(a^{-1})$, which are shown as red squares, 
have less $\beta$ dependence. 
This $Z_m(a^{-1})$ may well be expressed in a polynomial of $g^2$ with
$Z_m(a^{-1})\to 1$ in the continuum limit
\begin{equation}
 Z_m(a^{-1}) = 1 + \hat{c}_1 g^2 + \hat{c}_2 g^4 + \cdots.
\end{equation}
Therefore we adopt a fit which is an expansion in $\beta^{-1}$,
\begin{equation}
 Z_m(a^{-1}) = 1 + c_1 \beta^{-1} + c_2 \beta^{-2}.
\label{eq:Zm_param_binv2}
\end{equation}
The fit result is shown as a brown dashed line.
From this result one can obtain $Z_m(2\ \mbox{GeV})$ at arbitrary $\beta$ values
by applying the NNNLO running, which are shown as a green solid line
obtained with the $a(\beta)$ parameterization with Eq.~(\ref{eq:a_scale2}).

\begin{figure}
\begin{center}
 \includegraphics[width=7cm]{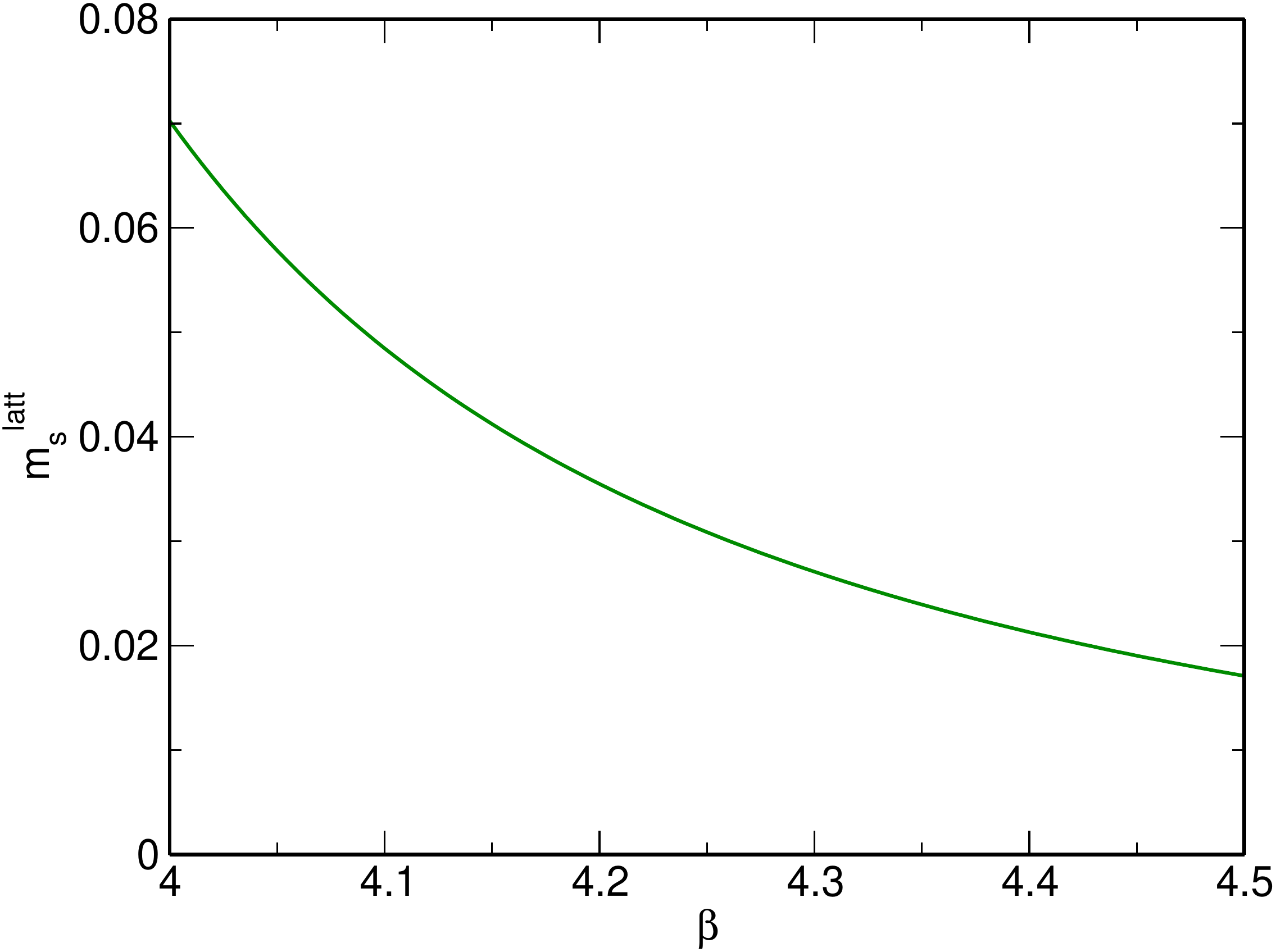}
 \includegraphics[width=7cm]{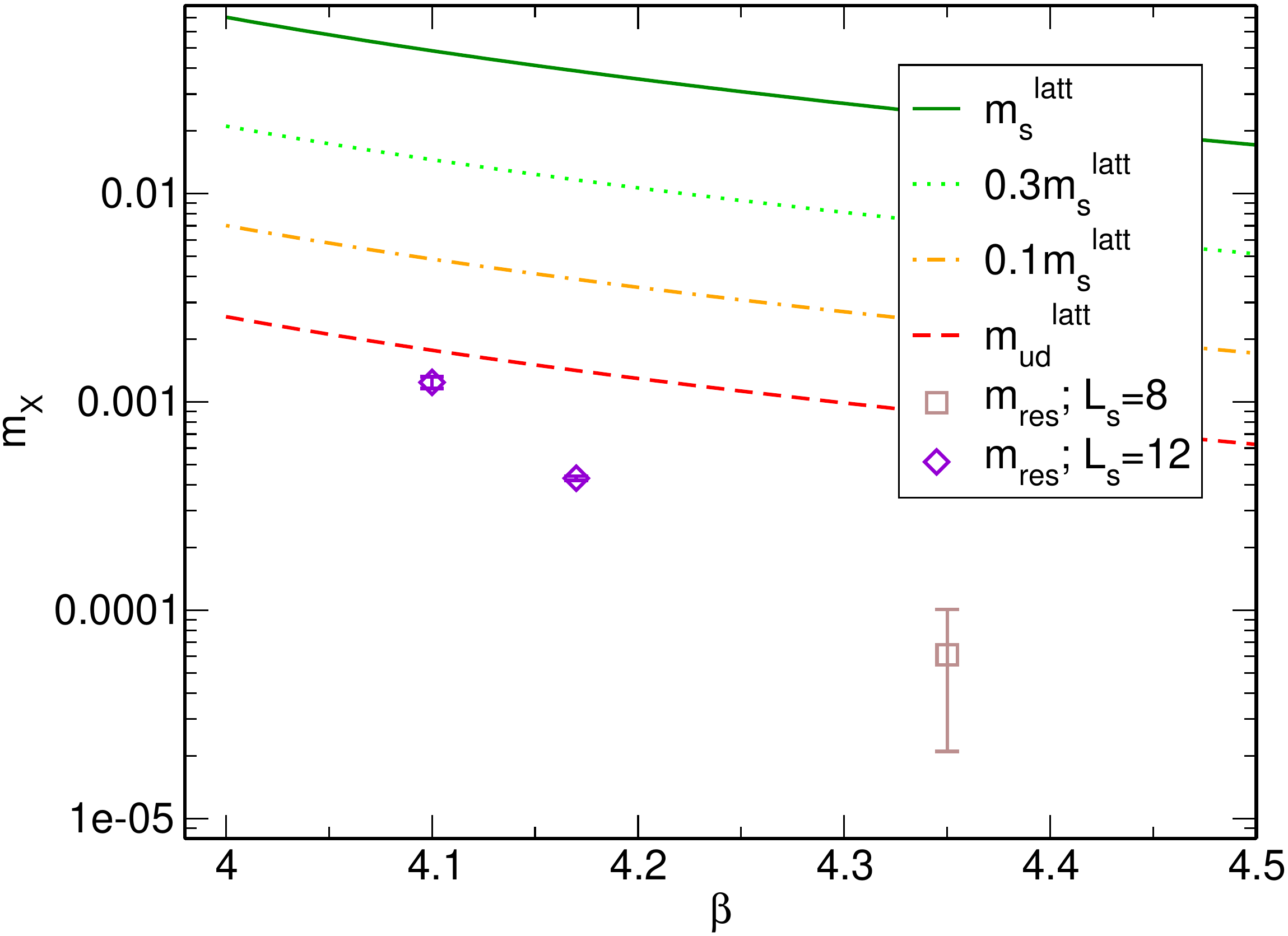}
\end{center}
 \caption{
 Line of constant physics for $m_s^{latt}$ (left).
 Line of constant physics for $m_s^{latt}$, $0.3 m_s^{latt}$, $0.1 m_s^{latt}$ 
 and $m_{ud}^{latt}$ in comparison with the residual quark mass (right).}
 \label{fig:ms_mX-beta}
\end{figure}

Using the parameterizations $a(\beta)$ and $Z_m(\beta)$ and the 
inputs Eq.~(\ref{eq:m_s}) $m_s^{latt}(\beta)$ is determined
and plotted in the left panel of Fig.~\ref{fig:ms_mX-beta}.
The right panel shows that with log-y scale, together with
$m_{ud}^{latt}(\beta)$ determined from $m_s^{latt}(\beta)$ with
Eq.~(\ref{eq:m_ud}), as well as $\alpha m_s^{latt}$ with $\alpha=0.3$ and $0.1$.
The residual quark mass computed on the three ensembles 
are also shown.
\footnote{The $m_{res}$ data are of $(\beta,m_s,m_{ud}) = 
(4.10,0.039,0.012),
(4.17,0.04,0.007),(
4.35,0.018,0.0042)$. The first one is newly computed and the others are from Ref.~\cite{JLQCD_Colquhoun:2021supp}.}

\section{Early results}

In this section some early and preliminary results are shown.
We use the scale setting and line of constant physics
obtained in the previous section to simulate finite temperature QCD
with fixed quark masses in physical unit and with varying temperature.

\begin{figure}
\begin{center}
 \includegraphics[width=7.3cm]{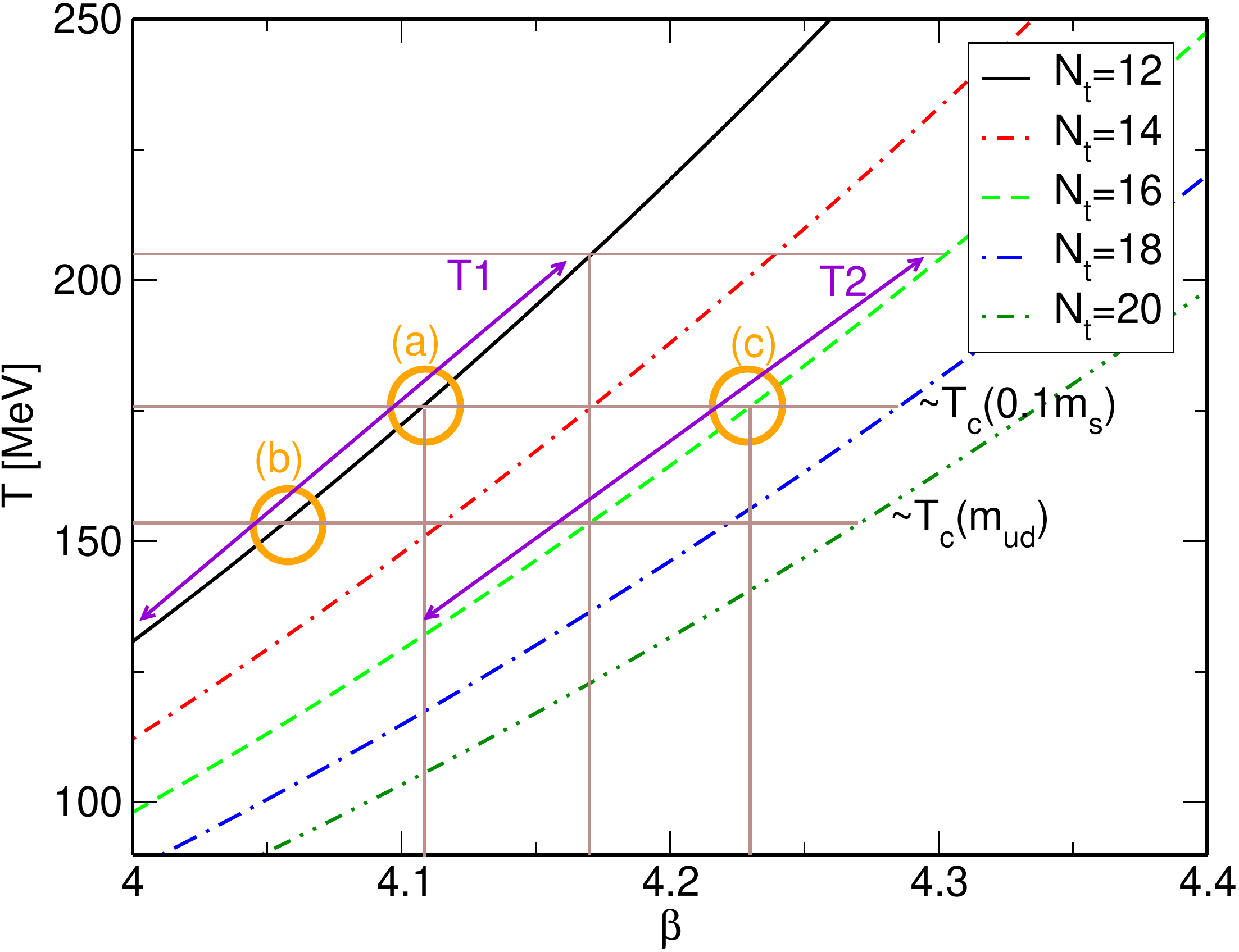}
\end{center}
 \caption{Temperature as a function of $\beta$ for different temporal lattice size $N_t$. Target $T-\beta$ trajectories are shown as T1 for $N_t=12$ and T2 for $N_t=16$. Expected transition regions are marked as (a), (c) for $m_l=0.1 m_s$, and (b) for physical $ud$ quark mass.}
 \label{fig:T-b}
\end{figure}

We adopt temporal size $N_t=12$ and $16$. 
The temperature - $\beta$ trajectories of interest are shown in
Fig.~\ref{fig:T-b}.

The strange quark mass is always tuned as physical (Eq.~(\ref{eq:m_s})).
The simulated average $ud$ quark mass is set as one tenth of the strange
for the moment ($m_l=m_s/10$).
Note that for now we neglect the effect of $m_{res}$.
The effect will be larger for coarser lattices and for smaller quark masses
with a fixed $L_s$.

\begin{figure}
\begin{center}
 \includegraphics[width=7.3cm]{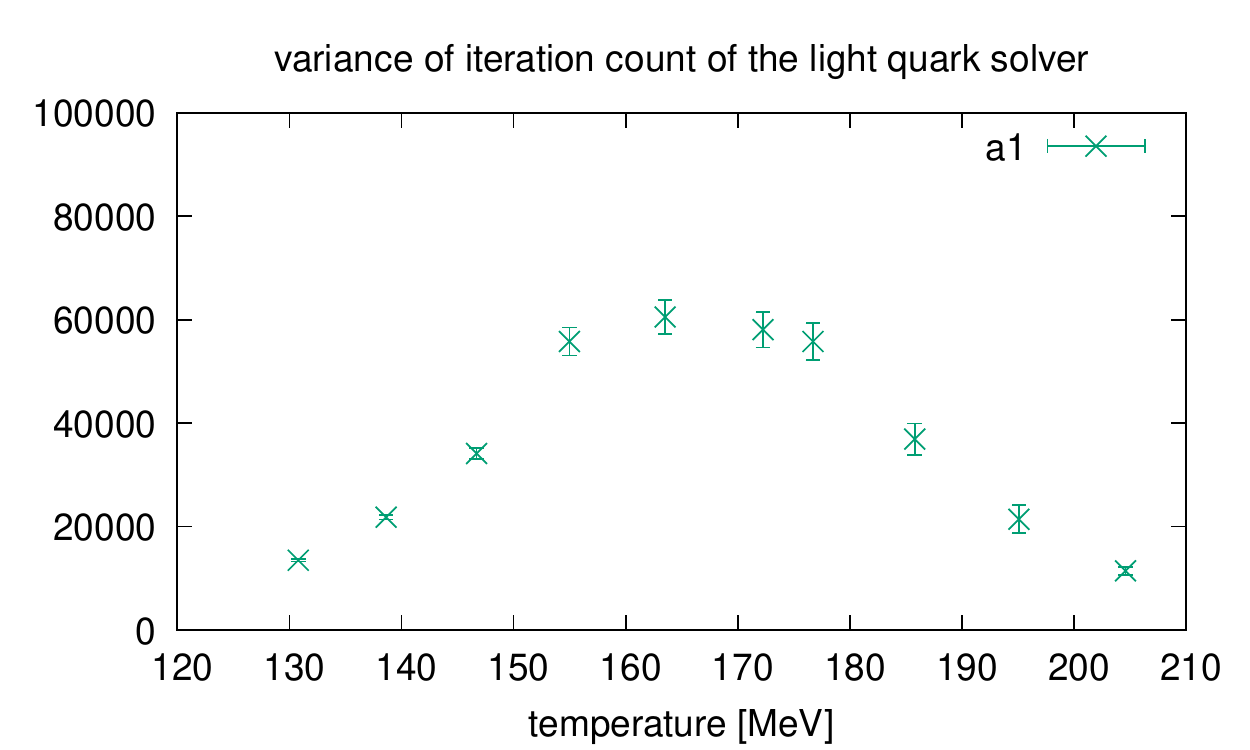}
\end{center}
 \caption{
 Variance of the iteration count of conjugate gradient for the light quark solver in the molecular dynamics at $N_t=12$ as a function of temperature.}
 \label{fig:CGvariance}
\end{figure}

Figure \ref{fig:CGvariance} shows the variance of the iteration count
of conjugate gradient for the light quark solver in the hybrid Monte Carlo
simulations at $N_t=12$. As a first step we have chosen rather small
aspect ratio of spatial and temporal lattice sizes as $N_s/N_t=2$.
It develops a peak around $T\simeq 170$ MeV,
which turns out to be consistent with the observation in a pilot study 
with fixed $\beta=4.17$ and varying $m_l$.
This quantity may be used as an indicator of the transition
as it should have a correlation with the fluctuation of physical quantities.
It is useful in pinning down the transition region prior to
performing various measurements on the configurations.

\begin{figure}
\begin{center}
 \includegraphics[width=7.3cm]{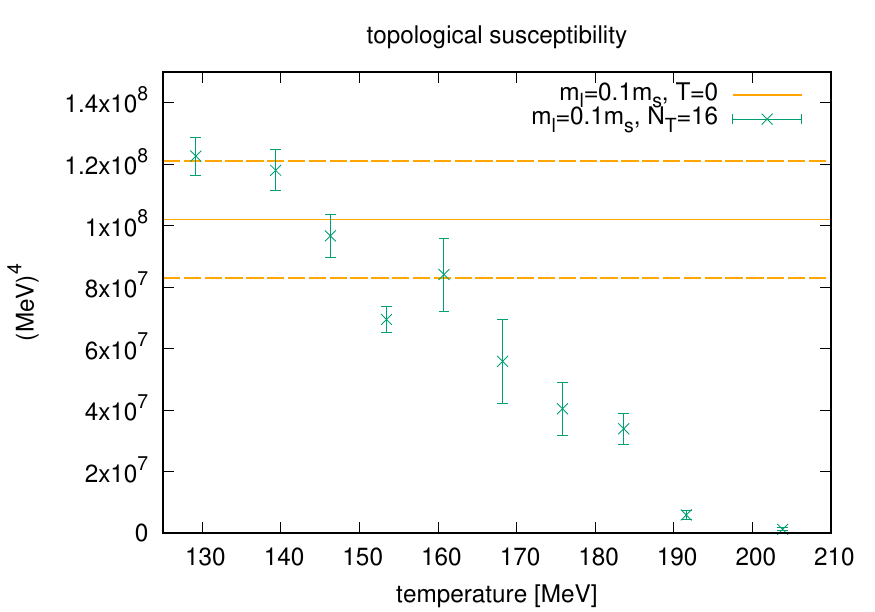}
\end{center}
 \caption{Topological susceptibility as a function of temperature
 for $N_t=16$ and with the line of constant physics, $m_l=m_s/10$.}
 \label{fig:chit}
\end{figure}

Figure \ref{fig:chit} plots the topological susceptibility at $N_t=16$
computed using a gluonic definition with a cooling using the Wilson flow.
Horizontal lines indicate the central value and error in the zero temperature
\cite{Aoki:2017paw}. At the low temperature region in this figure
the $m_{res}$ effect is getting sizable as this quantity is sensitive 
to the change of the quark mass. The correction due to the $m_{res}$ effect
is important in this region, which is yet to be determined.

\section{Summary and Outlook}
In this article a status report was provided on the systematic investigation 
of the
$2+1$-flavor finite-temperature QCD transition using M\"obius
domain-wall fermions at fine lattices up to $N_t=16$.
The line of constant physics has been determined, with which
a series of simulations have been performed aiming to get 
physics at the $ud$ quark mass being one tenth of the strange
((a) and (c) regions in Fig.~\ref{fig:T-b}).
Various measurements of the fermionic observables are now underway,
which will be used to determine the (pseudo) critical point and
related physics around that.
Understanding the size of $m_{res}$ around $\beta=4$,
which is the lower edge of our finite temperature simulations, 
and its correction to the physical quantities are important especially 
for the coarser lattice ($N_t=12$). With that further study with lowering
the light quark mass will be sought, using on-going simulations
in the range (b) in Fig.~\ref{fig:T-b}.

\section*{Acknowledgments}
This work is supported by MEXT as ``Program of Promoting Researches on the Supercomputer Fugaku'' (Simulation for basic science: from fundamental laws of particles to creation of nuclei) JPMXP1020200105, with HPCI project Nos.~hp200130 and hp210165 through that the following computers were used: 
the supercomputer Fugaku provided by the RIKEN Center for Computational Science, 
Oakforest-PACS provided by JCAHPC, and
Polaire and Grand Chariot at Hokkaido University.
This work is also supported in part by JSPS KAKENHI Grant No. 20H01907.
We acknowledge the use of Grid 
\footnote{https://github.com/paboyle/Grid}
and its extension
for A64FX processors \cite{Meyer:2019gbz} to generate the finite temperature
configurations used in this study. 
We thank N.~Meyer and T.~Wettig for discussions on the use of Grid for A64FX.
For the measurements for the new data in this study,
the following code packages were used:
BQCD      % Wilson flow for t0 on beta=4.1, T=0 lattices
  \footnote{https://www.rrz.uni-hamburg.de/services/hpc/bqcd.html},
Bridge++  % Wilosn flow and Qtop measurements for T>0
%  \footnote{https://bridge.kek.jp/Lattice-code/index_e.html}, and
  \footnote{https://bridge.kek.jp/Lattice-code/}, and
Hadrons   % new mres measurements
  \footnote{https://github.com/aportelli/Hadrons}.

\bibliographystyle{JHEP}
\bibliography{lattice2021_yaoki}
% \begin{thebibliography}{99}
% \bibitem{...}
% ....

% \end{thebibliography}

\end{document}